\documentstyle[11pt,epsf]{article}
\newcommand{\beq}{\begin{equation}}
\newcommand{\eeq}{\end{equation}}
\newcommand{\beqa}{\begin{eqnarray}}
\newcommand{\eeqa}{\end{eqnarray}}
\newcommand{\beqan}{\begin{eqnarray*}}
\newcommand{\eeqan}{\end{eqnarray*}}

\newcommand{\ben}{\begin{enumerate}}
\newcommand{\een}{\end{enumerate}}
\newcommand{\bfl}{\begin{flushleft}}
\newcommand{\efl}{\end{flushleft}}
\newcommand{\ba}{\begin{array}}
\newcommand{\ea}{\end{array}}
\newcommand{\btab}{\begin{tabular}}
\newcommand{\etab}{\end{tabular}}
\newcommand{\bit}{\begin{itemize}}
\newcommand{\eit}{\end{itemize}}

\def \g5 {\gamma_{5}}
\def \I {\hbox{ i}}
\def \W {\hbox{w}}
\def \D {\hbox{d}}
\def \Rf {\hbox{Fig}}
\def \L {\hbox {$\ell'$}}
 \def\psl{\not{\hbox{\kern-2.3pt $p$}}} 
\def\Psl{\not{\hbox{\kern-2.3pt $P$}}} 
\def\ksl{\not{\hbox{\kern-2.3pt $k$}}} 

\def\Gm{(1- \gamma_{5})}
\def\Gp{(1+ \gamma_{5})}

\def \E {\hbox{ e}}
\def\qsl{\not{\hbox{\kern-2.3pt $q$}}} 

\def \Z{\hbox{z}}

\newcommand{\prepr}[1] {\begin{flushright} {\bf #1} \end{flushright} \vskip
1.5cm}
\newcommand{\titul}[1] {\begin{center}{\Large {\bf #1 } } \end{center} \vskip 1.cm}

\newcounter{muni}

\begin{document}
\vspace{.1cm}
\hbadness=10000
\pagenumbering{arabic}
\begin{titlepage}
\topmargin 0mm
\textwidth 160mm
\textheight 220mm
\oddsidemargin 0mm
\evensidemargin 0mm

\prepr{Preprint PAR-LPTHE-99-07\\ February 1999 }

\titul{Massive Neutrino Decays.}

\vspace{5mm}
\begin{center}

{\bf  Pham Xuan Yem \footnote{\rm Postal address: LPTHE, 
  Tour 16, $1^{er}$ Etage, 
 4 Place Jussieu, F-75252 Paris CEDEX 05, France. \\
 \hspace{5mm} Electronic address : pham@lpthe.jussieu.fr }}
\end{center} 
\vspace{0.5mm}
 \begin{center}
 {\large \bf \it
  Laboratoire de Physique Th\'eorique et Hautes Energies, Paris \\
  CNRS, Universit\'e P. et M. Curie, Universit\'e D. Diderot} 
\end{center}

\thispagestyle{empty}
\begin{center}
\vspace{0.5mm}
\hspace{0.01cm} \large{} {\bf  \hspace{0.1mm} Abstract-}    
\vspace{0.5mm}
\normalsize
\end{center}
Lecture given at the Vth Viet Nam School of Physics, Hanoi 28 December 1998 - 9 January 1999. Neutrino physics is used  as an illustrative example for
an elementary introduction to the computational method of Feynman loop-diagrams and decay rates. 

If the neutrinos are as massive as recently reported by the Super-Kamiokande results, the heaviest neutrino $\nu_H$ would not be stable. Although chargeless, it could decay -- by quantum loop effect -- into a lighter neutrino $ \nu_L$ by emitting a photon: $\nu_H \to  \nu_L  + \gamma $. 
If kinematically possible, the 
$\nu_H \to \nu_L$ + e$^+$ + e$^-$ mode could occur at the tree diagram level and furthermore get enhanced, at one-loop radiative corrections, by a large logarithm of the electron mass acting as an infrared cutoff.

\vspace{10mm}
\end{titlepage}

\newpage

Evidence for the transmutation between the two neutrino species $\nu_\mu \leftrightarrow \nu_\tau$ is recently reported  by the Super-Kamiokande collaboration$^{(1)}$. As a consequence, neutrinos could have nondegenerate tiny masses, and mixing among different lepton families would occur, similarly to the Cabibbo--Kobayashi--Maskawa (CKM) flavor mixing in the quark sector. 
Let us start by assuming that the neutrino flavored states $\nu_{\rm e}$, $\nu_\mu$ and $\nu_\tau$ are linear combinations of the three neutrino mass eigenstates $\nu_1$, $\nu_2$ and $\nu_3$ of nonzero and nondegenerate masses $m_1$, $m_2$ and $m_3$ respectively. Thus
\begin{equation}
\pmatrix{ \nu_{\rm e}\cr \nu_\mu\cr \nu_\tau\cr}
=\pmatrix { U_{{\rm e}1} & U_{{\rm e}2} & U_{{\rm e}3}\cr
                 U_{\mu 1} & U_{\mu 2 } & U_{\mu 3}\cr
                 U_{\tau 1}  & U_{\tau 2} & U_{\tau 3}\cr}
               \pmatrix {\nu_1\cr \nu_2 \cr \nu_3\cr} 
\equiv \; {\cal U}_{\rm lep}\;\pmatrix {\nu_1\cr \nu_2 \cr \nu_3\cr } , 
\label{eq:1} 
\end{equation}
where the $3\times 3$ matrix $\;{\cal U}_{\rm lep}\;$ is unitary. Neutrino oscillation measurements give constraints usually plotted in the 
$(\sin 2\theta_{i j}$,  $\Delta m_{i j}^2= |m^2_i-m_j^2|)$ plane, where
 $\theta_{i j}$ is one of the three Euler angles of the rotation matrix $\;{\cal U}_{\rm lep}$.  The effective weak interactions of leptons can now be written as
$${\cal L}_{\rm eff} ={{\rm G}_{\rm F} \over \sqrt{2}} L^{\dagger}_\lambda L_\lambda \;,$$ 
where the charged current $L_\lambda$ is
$$L_\lambda = \sum_{\ell}\sum_{i =1}^{3}  U_{\ell i}\overline{\nu}_i\gamma_{\lambda}(1-\gamma_5) \ell\;.$$
Here $\ell$ stands for e$^-$, $\mu^-$, $\tau^-$ and $\nu_i$ (with $i=1,2,3$) are the three neutrino mass eigenstates. Unitarity of ${\cal U}_{\rm lep}$ 
implies that for any fixed $\ell$, one has $\sum_{i}\vert U_{\ell i}\vert^2 =1$,  or $\sum_{i}U_{\ell i} U^*_{\ell' i} =\delta_{\ell \ell'}$ and $\sum_\ell U_{\ell i} U^*_{\ell j} =\delta_{i j}$. This current $L_\lambda$ tells us for instance that the neutrino $\nu_\mu$ operationally defined to be the invisible particle missing in the decay $\pi^+ \to \mu^{+} + \nu_{\mu}$ is initially a superposition of $\nu_1$, $\nu_2$ and $\nu_3$, in the same way as the K$^0$ meson produced by strong interaction is initially a superposition of the mass eigenstates K$^0_{\rm L}$ and K$^0_{\rm S}$ with masses $m_{\rm L} \not= m_{\rm S}$. The nondegenerate masses give rise to the oscillation phenomena in both neutral K mesons and neutrinos.
 
 Moreover, although the neutrinos are chargeless, a heavy neutrino $\nu_H$ could decay into a lighter neutrino $\nu_L$ by emitting 
a photon; this  decay is entirely due to  
quantum loop effects. If kinematically possible, i.e. if the  $\nu_H$--$\nu_L$ mass difference is larger than twice the electron mass ($\approx 1$ MeV), then the mode $\nu_H \to \nu_L + \E^+ +\E^-$ largely dominates, because it is governed by a tree diagram and enhanced by  radiative corrections, as we will see.
\newpage
\hskip-7mm{\bf 1- One-loop effective $\nu_i$-$\nu_j$-$\gamma$ vertex $\Gamma^\mu$}

In gauge theories with spontaneous symmetry breaking, by the power counting arguments, renormalizability of electroweak interactions is not manifest in the unitary U gauge$^{2,3}$ where only physical gauge  bosons  W, Z are involved. This is because of the bad high-energy behavior of massive W, Z propagators [$D(k)\to 1/M_{\W,\Z }^2$ for $k\to \infty$
].  With the U gauge, except for lowest orders (tree diagrams),  
it is  practically impossible to perform higher order calculations, because 
the $k$ integrations in loop diagrams cause an avalanche of uncontrollable quadratic divergences. 

On the other hand, the renormalizable gauge$^{2,3}$ (conventionally called $R_\xi$) is particularly convenient for loop calculations. As shown by 't Hooft, the key is to choose a gauge parameter $\xi$ with which the gauge boson  propagator has a mild high-energy behavior ($D(k)\to 1/k^2$  for $k\to \infty$), at the expense of introducing fictious particles, the "would be" Goldstone bosons $\Phi$ (those absorbed by the weak gauge bosons to render them massive by the 
Higgs mechanism). Feynman rules for the W and $\Phi$ propagators, as well as for some vertices in the  $R_\xi$ gauge are  given in the appendix, from which we compute  loop amplitudes.

In  $R_\xi$ at one loop level, six Feynman diagrams contribute to the process  $\nu_H (P) \to \nu_L(p) + \gamma(q)$, where the photon can be real ($q^2=0$) or virtual ($q^2\neq 0$); the latter is necessary when we consider the radiative corrections to the  $\nu_H \to \nu_L + \E^+ +\E^-$ tree diagram.  These six diagrams can be grouped into two sets: four in Figs. 1a$-$1d and two in Fig.2a$-$2b. Every diagram is gauge-dependent through a $\xi$ parameter, however the $\xi$-dependences are canceled out separately for each group, i.e. the sum of the four diagrams in Fig.1 is $\xi$-independent; the same occurs for the sum of the two diagrams in Fig.2. Consequently, the final result is {\it gauge-independent}, 
as it should be 
for any physical process. In the U gauge corresponding to $\xi=\infty$ , only two diagrams (Fig.1a and Fig.2a) contribute; the price to pay is the bad high-energy behavior of the W boson propagator which renders the loop integration particularly difficult.

$\bullet$ Let us explicitly compute, as an example, the simplest amplitude ${\cal A}_{1b}$ associated with the diagram of Fig.1b in the general $R_\xi$ gauge for which Feynman rules give
\begin{equation}
\I \; {\cal A}_{1b} = (-\I e)\left({ \I g \over 2 \sqrt{2}  }\right)^2 
\sum_\ell U_{L\ell} U^*_{H\ell}\;\overline{u}(p) \,\Gamma_{1b}^\mu(\ell)  \,u(P)\;\varepsilon^*_\mu (q)\;,
\end{equation}
\begin{equation} 
\Gamma_{1b}^\mu(\ell)  = \int {\D^4 k\over (2\pi)^4} 
{[m\Gm -m_\ell\Gp] [\I (\ksl +m_\ell)] [M\Gp -m_\ell\Gm](P+p-2k)^\mu
 \over M_{\W}^2 \;[(k-p)^2- \xi M_W^2] [(P-k)^2 -\xi M_W^2] (k^2-m_\ell^2)} \;,
\end{equation}
where $m$ and $M$ are respectively the light $\nu_L$ and the heavy $\nu_H$ neutrino masses, $m_{\ell}$ ($\ell=1,2,3$) stand for e$^-$, $\mu^-$ and $\tau^-$ masses, and $\varepsilon^*_\mu (q)$ denotes the photon polarization four-vector. $\Gamma_{1b}^\mu(\ell)$ is the contribution of Fig.1b to the effective neutrinos-photon vertex $\Gamma^\mu$. To simplify the computation we will only keep $M \neq 0$, and neglect $m$ in the following, since $m$ is smaller than  $M, m_\ell$ or $ M_{\W} $. Inserting $\Gamma_{1b}^\mu (\ell)$ between $\overline{u}(p)$ and $u(P)$, making use of Dirac equations for these spinors and adopting the standard Feynman parameterization$^2$ for the denominator of (3):
$$\hskip-2mm{1\over [(k-p)^2- \xi M_W^2] [(P-k)^2 -\xi M_W^2] [k^2-m_\ell^2]} =\int_0^1 \D x \int_0^{1-x} {2\D y\over [k^2-2k.(pz +Py) -(\xi M_W^2(1-x) +
m_\ell^2 x)]^3} $$%
with $z=1-x-y$.  When we perform the $\D^4 k$ integration of (3), the quadratic term in  $\,k\,$ of the numerator  yields a logarithmic ultraviolet (UV) 
divergence. To handle this UV, we adopt$^2$ the 't Hooft--Veltman  dimensional regularization by replacing the space-time $4$ dimensions with $ n +2 \epsilon$ ($4\leftrightarrow n+2\epsilon$), and the UV is symbolized by the singular Euler function $\Gamma(\epsilon) \approx 1/\varepsilon$ for $\epsilon \to 0$. We will show that these $\Gamma(\epsilon)$ either  mutually cancel out among the six diagrams, or identically vanish due to $\sum_{\ell} U_{L \ell } U^*_{H \ell}  =0$  reflecting the Glashow--Iliopoulos--Maiani (GIM) mechanism. The final result which must be finite is  obtained by putting $\epsilon =0$ at the end. The presence of $\gamma_5$ in $n$ dimensions does not cause any ambiguity because we do not compute the traces of 
Dirac matrices in our $\D^n k$ integration here. (About the problem of $\gamma_5$ in $n$ dimensions, see for instance reference 2). 

We get after the $\D^n k$ integration of (3):
\begin{equation} 
\Gamma_{1b}^\mu(\ell)  = {1\over 8\pi^2}\Gp \int_0^1 \D x \int_0^{1-x} \D y { {\cal N}_{b}^\mu \over {\cal D}_1 (\ell)}\;,
\end{equation}
where
\begin{equation} 
{\cal D}_1 (\ell) =M_{\W}^2 [\xi (1-x) +r_\ell x] -M^2 x y -q^2 \,y(1-x-y)\;\;,\; r_\ell=
 {m_\ell^2\over M_{\W}^2}\;,
\end{equation}
 \begin{equation} {\cal N}_{b}^\mu  = r_\ell \Bigl\{M (1-y) [(2y-1)P^\mu +(1- 2x - 2y)p^\mu] + {\cal D}_1 (\ell)\,\left\{\Gamma(\epsilon) -\log [{\cal D}_1 (\ell)/M_W^2] \right\} \gamma^\mu \Bigr\} \;.
\end{equation}

Let us first discuss the question of UV divergences in the six amplitudes.
As can be seen  in (17) and (24), the divergences $\Gamma(\varepsilon)$ of diagrams 1a and 2a are multiplied by $\ell$-{\it independent} coefficients, respectively $+6$ and $-2$. Because of this fact, these UV become harmless when the three internal charged lepton  contributions are summed up, 
 due to  the unitarity of $\;{\cal U}_{\rm lep}\;$ which tells us that $\sum_\ell U_{L\ell} U^*_{H\ell}\;\Gamma(\epsilon) =  \Gamma(\epsilon) \sum_\ell U_{L\ell}U^*_{H\ell} = 0$. This is the essence of the GIM mechanism$^2$.
On the other hand, as can be seen explicitly in (9), the coefficient $r_\ell$ of  $\Gamma(\varepsilon)$ in $\Gamma_{1b}^\mu(\ell)$ is $\ell$-{\it dependent}, so  $\sum_\ell U_{L\ell} U^*_{H\ell} \; r_\ell \,\Gamma(\epsilon)   \neq 0$. This UV will be exactly canceled by that in Fig.2b, as given by (27).  Finally, the amplitudes of diagrams 1c and  1d are ultraviolet convergent. 

The  $\log [{\cal D}_1 (\ell)/M_{\W}^2]$ term in (6) is the finite part extracted from 
$${\cal D}^{-\epsilon}_1 (\ell) \Gamma(\epsilon) = [1-\epsilon 
\,\log({\cal D}_1 (\ell)/M_{\W}^2)] \Gamma(\epsilon) +{\cal O}(\epsilon)\;.$$%
The first term of  $ {\cal N}_{b}^\mu $ on the right hand side of (6) can be rewritten as
\begin{equation}
M (1-y)\left[- 2 x  P^\mu + (2x+2y-1)q^\mu\right]\;.
\end{equation}
By translational invariance,  the neutrinos-photon vertex $\Gamma^\mu$ depends only on the four-momentum transfer $q^\mu$ and not on
 $P^\mu$, the latter may be written as a combination of three independent vectors $\I \sigma^{\mu \nu} q_\nu \,$, $\;q^\mu,$ and $\gamma^\mu$ by using the following relation valid for $m=0$, 
\begin{equation}
2 \,\overline{u}(p) (1+\gamma_5) P^\mu u(P)= \overline{u}(p) (1+\gamma_5)[\I \sigma^{\mu \nu} q_\nu  +M \gamma^\mu + q^\mu ]u(P)\;.
\end{equation}
 Putting altogether (4)$\cdots$(8), the ratio $ {\cal N}_{b}^\mu/ {\cal D}_1 (\ell) $ in (4)
 can be rewritten as 
\begin{equation}
{{\cal N}_{b}^\mu \over {\cal D}_1 (\ell)}  =r_\ell \Bigl \{ { \I M\sigma^{\mu\nu} q_\nu x(y-1)\over {\cal D}_1 (\ell)}  + 
{\qsl q^\mu \over {\cal D}_1 (\ell)}  (y-1)(1-x-2y) + {M^2\gamma^\mu\over {\cal D}_1 (\ell)} x(y-1) + \Bigl [\Gamma(\varepsilon)-
\log [{\cal D}_1 (\ell)/ M_W^2]  \Bigr] \gamma^\mu \Bigr \}\,. \end{equation}
For real photon emission  $\nu_H\to \nu_L +\gamma$, only the $\sigma^{\mu\nu} q_\nu$ term in (9) contributes to the decay amplitude$^{3,4}$; this property is due to the conservation of the electromagnetic current, i.e. $\partial{_\mu} J_{em}^\mu =0$. Indeed, the $\nu_H\to \nu_L +\gamma$ amplitude is written as  $ < \nu_L(p)| J_{em}^\mu |\nu_H (P)>\varepsilon_\mu^*(q)$ and the matrix element of the current $J_{em}^\mu$  has the most general Lorentz covariant form 
$$ < \nu_L(p)| J_{em}^\mu|\nu_H (P)> = \overline{u}(p)\Bigl[ (a+b\gamma_5) \I\sigma^{\mu\nu} q_\nu + (c +d\gamma_5)\gamma^\mu + (e + f \gamma_5)q^\mu \Bigr]u(P)\;.$$%
The condition $q_\mu < \nu_L(p)| J_{em}^\mu|\nu_H (P)>=0$  implies that only the magnetic form factors $a, b$  survive, the $c,d$ terms must vanish using the Dirac equation applied to $\overline{u}(p)$, $u(P)$, and finally the $e,f$ terms proportional to $q^\mu$ do not contribute when multiplied by $\varepsilon_\mu^*(q)$. 

Note that for the  virtual photon case as in $\nu_H(P)\to \nu_L(p) +$ e$^+(k_{+}) $ + 
e$^-(k_-)$ of Fig.4, the $q^\mu =( k_- + k_{+})^\mu$ term  is also irrelevant because it  vanishes when contracted with  the conserved vector current $ \overline{u}(k_-)\gamma_\mu v(k_{+}) $ of the  electron-pair .

To obtain the $\nu_H\to \nu_L +\gamma$ decay amplitude ${\cal A}_{1b}$ from Fig.1b according to (2)--(6), we  perform the $x, y$ integrations of the first term $\;[x(y-1)/{\cal D}_1 (\ell)]\I M\sigma^{\mu\nu} q_\nu $ in (9).  For that,  we  will neglect $M^2xy  \ll M_{\W}^2$ in ${\cal D}_1 (\ell)$ to simplify the computation, and put $q^2=0$ since we are dealing with a real photon. Thus (2) becomes:
\begin{equation}
{\cal A}_{1b}= {\cal A}_0 \sum_\ell U_{H\ell} U^*_{L\ell} F_{1b}(\ell)\;,
\end{equation}
where
\begin{equation}
 {\cal A}_0 ={G_F\over \sqrt{2}} {e\over 8\pi^2}\overline{u}(p)[M \Gp]
 \I \sigma^{\mu \nu} q_\nu u(P)\;\;\varepsilon^{*}_\mu(q) 
\end{equation}
\begin{equation}
\hskip-8mmF_{1b}(\ell) =  { \xi r_\ell^2 (r_\ell-2\xi) \log (r_\ell /\xi)\over 2(r_\ell-\xi)^4} +r_\ell \left[-\,{1\over 3 (r_\ell-\xi) } - \,{\xi\over 4 (r_\ell-\xi)^2} +{\xi^2\over 2(r_\ell-\xi)^3} \right]\;, 
\end{equation}
with the Fermi constant $G_F=g^2 / (4 \sqrt{2} M_W^2)$. If $m$ is not neglected, the $[M \Gp]$ term in (11) would be replaced by
$[M \Gp +m \Gm]$. If we keep $M^2xy$ in ${\cal D}_1 (\ell)$, we still obtain an explicit analytic form for $F_{1b}(\ell)$; the result is complicated and not illuminating however. The exact formula (12) is in agreement with  similar calculations$^3$  for $\mu^- \to $ e$^- +\gamma$  in the limit $r_\ell \to 0$, where only the linear term in $r_\ell$ was kept and the logarithmic term neglected. In the same reference 3, the $\xi$-dependences  of the four diagrams similar to Fig.1a--d are explicitly shown to mutually canceled out, leaving the final result $\xi$-independent. 
In the following, we will compute  the five other amplitudes in the Feynman--'t Hooft gauge ($\xi=1$), for which (12) becomes
\begin{equation} 
\hskip-8mmF_{1b}(\ell) =  {r_\ell^2 (r_\ell-2) \log r_\ell\over 2(r_\ell- 1)^4} +r_\ell \left[-\,{1\over 3 (r_\ell-1) } - \,{1\over 4 (r_\ell-1)^2} +{1\over 2(r_\ell-1)^3} \right]\;. 
\end{equation}
The singularities of $F_{1b}(\ell)$ at $r_\ell=1$ are only apparent, in fact $F_{1b}(\ell) =-1/8$ for $r_\ell=1$. 

$\bullet$ By the same method just outlined, the ${\cal A}_{1a}$ amplitude of Fig.1a in the $\xi=1$ gauge is given by 
\begin{equation}
\I\; {\cal A}_{1a} = (\I e)\,\left({ -\I g\over 2\sqrt{2}}\right)^2 \sum_\ell U_{L\ell}U^*_{H\ell} \overline{u}(p)\Gamma^\mu_{1a}(\ell)\, u(P) \varepsilon^*_\mu (q)
\end{equation}
$$ \Gamma^\mu_{1a}(\ell) = \int {\D^4 k\over (2\pi)^4} {\gamma_{\rho} \Gm [\I(\ksl +m_\ell)] \gamma_\sigma \Gm X^{\mu \rho \sigma}
 \over [(k-p)^2- M_W^2] [(P-k)^2 - M_W^2] (k^2-m_\ell^2)} \;\;$$
\begin{equation}
X^{\mu \rho \sigma} = (k+p-2P)^\rho \, g^{\mu \sigma} + (k+P-2p)^\sigma \, g^{\rho \mu} + (P+p-2k)^\mu \, g^{\rho \sigma} \;.
\end{equation}
We get
\begin{equation}
 \Gamma^\mu_{1a} = { 1 \over 8 \pi^2} \Gp  \int_0^1 \D x \int_0^{1-x} \D y  
{{\cal N}_{1a}^\mu \over {\cal D}_1(\xi=1, x)}\;\ 
\end{equation}
$$
{\cal N}_{1a}^\mu  =\Bigl\{[2(1-x)(1-y)+y] M^2 - 2 [(1-x)(1-y) +y^2] 
\, q^2
+6 \,{\cal D}^{1-\epsilon}_1 (\ell) \Gamma(\epsilon) \Bigr\} \gamma^\mu $$%
 \begin{equation} + 2M  \left \{y (1-2y) P^\mu +[2y^2 -(1-x)(1+2y)] p^\mu \right\}\end{equation}
The first term of ${\cal N}_{1a}^\mu$ proportional to $\gamma^\mu $ does not contribute to the real photon process $\nu_H\to \nu_L +\gamma$. The second term, coefficient 
of $2M $, may be rewritten as
$$y (1-2y) P^\mu +[2y^2 -(1-x)(1+2y)] p^\mu =   [y-(1-x)(1+2y)]\,P^\mu  +C q^\mu \;,$$%
and as before the  $P^\mu$ term is converted into  form (8). We get for (16)  
$$\hskip-4.2cm{{\cal N}_{1a}^\mu \over {\cal D}_1(\ell)} ={ \I M\sigma^{\mu\nu} q_\nu\over {\cal D}_1 (\ell)} [x-1 -y(1-2x)]  + 
{\qsl q^\mu \over {\cal D}_1 (\ell)} [1-x+3y-2xy-4y^2]  $$%
\begin{equation}+ {M^2\gamma^\mu\over {\cal D}_1 (\ell)} [1-x-2y(1-2x)] - {2q^2\gamma^\mu\over {\cal D}_1 (\ell)} [y^2+(1-x)(1-y)] 
+6 \Bigl\{\Gamma(\varepsilon)-
\log [{\cal D}_1 (\ell)/ M_W^2]  \Bigr\} \gamma^\mu \;. 
\end{equation}
After the $x, y$ integrations of the first term in (18) relevant for the real photon case, we obtain the contribution from Fig.1a written in the form (10)--(12) with $F_{1b}(\ell)$ replaced by $F_{1a}(\ell)$, the latter is given  by 
\begin{equation}
F_{1a}(\ell) = {r_\ell^2 (1-3r_\ell) \log r_\ell \over 2(r_\ell-1)^4} +r_\ell \left[{7\over 12 (r_\ell-1) } +{2\over (r_\ell-1)^2} +{1\over (r_\ell-1)^3} \right] -{7\over 12}
\end{equation}
The singularities of $F_{1a}(\ell)$ at $r_\ell=1$ are only apparent, actually $F_{1a}(\ell)$ is equal 
to $-5/12$ for $r_\ell =1$. 

The remaining amplitudes derived from the diagrams of Fig.1c,d and Fig.2a,b are
 given below:
\begin{equation} \hskip-2cm  \bullet \Rf .1c :\hskip1cm {{\cal N}_{1c}^\mu \over {\cal D}_1(\ell)}  ={ \I M\sigma^{\mu\nu} q_\nu\over {\cal D}_1 (\ell)} (x+y -1)  + 
{\qsl q^\mu \over {\cal D}_1 (\ell)} (1-x-y) +{M^2\gamma^\mu\over {\cal D}_1 (\ell)} (x-1) +{m_\ell^2\gamma^\mu\over {\cal D}_1 (\ell)} \;,
 \end{equation}
which gives after the $x, y$ integrations
\begin{equation}
\hskip-3.8cmF_{1c}(\ell)   = {-r_\ell^2  \log r_\ell \over 2(r_\ell-1)^3} +r_\ell \left[{1\over 4 (r_\ell-1) } +{1\over 2 (r_\ell-1)^2}  \right] -{1\over 4}\;. 
\end{equation} 
\begin{equation}\hskip-12 cm \bullet\Rf .1d : \hskip 1cm {{\cal N}_{1d}^\mu \over {\cal D}_1(\ell)}  ={m_\ell^2\gamma^\mu\over {\cal D}_1 (\ell)} \;,
 \end{equation}
thus
\begin{equation}
\hskip-11cm F_{1d}(\ell)  = 0\;.
\end{equation}
$$ \hskip-4.3cm \bullet \Rf .2a:\hskip1cm {{\cal N}_{2a}^\mu \over {\cal D}_2(\ell)}  = { 2 \I M\sigma^{\mu\nu} q_\nu\over {\cal D}_2 (\ell)} x(y-1)  + 
 {2\qsl q^\mu \over {\cal D}_2 (\ell)} (1-y)(x+2y) - {2m_\ell^2\gamma^\mu\over {\cal D}_2 (\ell)}  $$%
\begin{equation}  + {2q^2 \gamma^\mu \over {\cal D}_2 (\ell)} (y-1)(x+y)- 2\Bigl\{\Gamma(\varepsilon)-
\log [{\cal D}_2 (\ell)/ M_W^2]  \Bigr\} \gamma^\mu \;,
 \end{equation}
where \begin{equation}  \hskip-3.2cm {\cal D}_2 (\ell)=M_{\W}^2 x + m_\ell^2 (1-x) 
-M^2 x y -q^2 \,y(1-x-y)\;.\end{equation}
We get
\begin{equation}
\hskip3.5mm F_{2a}(\ell)   =   {r_\ell (2r_\ell-1) \log r_\ell \over (r_\ell-1)^4} +
r_\ell \left[{2\over 3 (r_\ell-1) } -{3\over 2 (r_\ell-1)^2}  -{1\over 
(r_\ell-1)^3} \right] - {2\over 3} \;.
\end{equation} 
Finally,
$$ \hskip-2.5cm \bullet \Rf .2b:\hskip1cm {{\cal N}_{2b}^\mu \over {\cal D}_2(\ell)}  =r_\ell \Bigl\{ {  \I M\sigma^{\mu\nu} q_\nu\over {\cal D}_2 (\ell)} 
[x(1+y)-1]  + 
 {\qsl q^\mu \over {\cal D}_2 (\ell)} [1-x(1+y)- 2y^2] - {m_\ell^2\gamma^\mu\over {\cal D}_2 (\ell)}   $$%
\begin{equation}  +{M^2\gamma^\mu\over {\cal D}_2 (\ell)} x  + {q^2 \gamma^\mu \over {\cal D}_2 (\ell)} y(x+y-1) -\Bigl\{\Gamma(\varepsilon)-
\log [{\cal D}_2 (\ell)/ M_W^2]  \Bigr\} \gamma^\mu \; \Bigr\},
 \end{equation}
from which
\begin{equation}
F_{2b}(\ell)   =    {r_\ell(2-r_\ell) \log r_\ell \over 2(r_\ell-1)^4} +r_\ell \left[{-5\over 12 (r_\ell-1) } +{3\over 4 (r_\ell-1)^2}  
-{1\over 2(r_\ell-1)^3} \right]\;.
\end{equation} 
 The constants $(-7/12), (-1/4), (-2/3$) respectively in (19), (21) and (26)  being $\ell$-independent do not contribute to the decay amplitude when  summed over $\ell$, due to  $\sum_\ell U_{H \ell}^* U_{L\ell}=0$.
 The sum of the six terms $\sum_{\ell}U_{H\ell} U^*_{L\ell} [ F_{1.a\cdots d}(\ell) +F_{2.a,b} (\ell)]$ yields the final result for the $\nu_{H}\to \nu_{L} +\gamma$ decay amplitude
\begin{equation}
{\cal A}(\nu_{H}\to \nu_{L} +\gamma) = {3{\cal A}_0\over 4}   \sum_\ell {U_{H\ell} U^*_{L\ell}
\;r_\ell\over (1-r_\ell)^3}\left[1-r^2_\ell +2r_\ell\log r_\ell\right]\;,
\end{equation} 
where ${\cal A}_0$ is defined in (11). Our result (29) agrees with formula (10.28) for the function $f(r)$ in reference 4, where the three irrelevant constants mentioned above are kept. 
We get
\begin{equation}
\Gamma_0\equiv \Gamma(\nu_{H}\to \nu_{L} +\gamma)= {G_F^2 M^5\over 192 \pi^3} 
\left({ 27\alpha \over 32 \pi}\right) \left|
  \sum_\ell {U_{H\ell} U^*_{L\ell}
\;r_\ell\over (1-r_\ell)^3}\left[1-r^2_\ell +2r_\ell\log r_\ell\right]\right|^2\;.
\end{equation} 
We assume for ${\cal U}_{\rm lep}$ the following form$^5$ :
\begin{equation}
{\cal U}_{\rm lep} = \pmatrix{ \cos\theta_{12} & -\sin\theta_{12}& 0 \cr 
{1\over \sqrt{2}}\sin\theta_{12} & {1\over \sqrt{2}}\cos\theta_{12}  & 
{-1\over \sqrt{2}}\cr
        {1\over \sqrt{2}}\sin\theta_{12}&{1\over \sqrt{2}}\cos\theta_{12}  & 
{1\over \sqrt{2}}\cr}\,.        
\label{eq:4} 
\end{equation}
The mixing angle $\theta_{23}\approx 45^0$ is suggested by the Super-Kamiokande data and
the $\theta_{13}\approx 0^0$ comes from the Chooz data$^{1,5}$ which give $\theta_{13} \leq 13^0$, whereas $\theta_{12}$ is arbitrary.
Although  $\theta_{12}$ is 
likely small $\approx 0^0$, the maximal mixing $\theta_{12}\approx 45^0$ could be  possible which would allow $\nu_{\rm e}\leftrightarrow \nu_\mu$  oscillations (as suggested by the LSND experiment$^{1,5}$).  
Taking $\theta_{12}$ in the range 
$0^0$--$45^0$, and $M=5\times 10^{-2}$ eV, the decay rate $\Gamma(\nu_{H}\to \nu_{L} +\gamma)$  is found to be $\approx 10^{-44}$/year. 

\hskip-7mm{\bf 2- The mode $\nu_H \to \nu_L +$ e$^+$ + e$^-$}

If kinematically allowed, i.e. if the mass difference $M$--$m$ between the two neutrinos is larger than twice the electron mass, the decay $\nu_H (P)\to \nu_L(p) +$ e$^+(k_{+})$ + e$^-(k_{-})$ is possible, and governed by the tree diagram in Fig.3. The corresponding amplitude is
\begin{equation}
{\cal A}_{tree} = {G_F \over \sqrt{2}} U^*_{H e} U_{L e}\; \overline{u}(k_{-}) \gamma^\mu \Gm  
u(P) \;\overline{u} (p) \gamma_\mu \Gm v(k_{+}) \;.
\end{equation}
Using Fierz recombination$^2$ and $U^*_{H e} U_{L e} =-\sum_{\L= \mu, \tau} U^*_{H \L} U_{L \L}$, this amplitude can be rewritten as
\begin{equation}
{\cal A}_{tree} = (-1)^2 {G_F \over \sqrt{2}} \sum_{\L= \mu, \tau} U^*_{H \L} U_{L \L}\; \overline{u} (p) \gamma^\mu \Gm  
u(P) \;\overline{u}(k_{-}) \gamma_\mu \Gm v(k_{+})\;.
\end{equation}
Electromagnetic radiative corrections to ${\cal A}_{tree}$ come from the six diagrams previously considered; the emitted photon is now virtual and creates an electron pair (Fig.4).
 
A careful examination of all the terms  in (9), (18), (20), 
(22), (24) and (27) of the six vertices 
$\Gamma^\mu_{1a-d}(\ell),\Gamma^\mu_{2a,b}(\ell)$ shows 
that after the $x, y$ integrations, the largely dominant contribution comes from
Fig.2a with the  $q^2$ term  in (24) which exhibits an  infrared-like divergence
$\log\,r_\ell \to\infty$  for 
$r_\ell \to 0$ . 
 We can track down this  divergent behavior by examining the integration limits $x=0$ and $x=1$ of the denominators ${\cal D}_{1,2}(\ell)$. Infrared-like divergence  occurs if the 
numerators ${\cal N}^\mu_2(\ell)$ lack an $x$ 
term to cancel the $x=0$ integration limit of the $x\,M_{\W}^2$ 
term in the denominator ${\cal D}_2(\ell)$ in (25). This happens with the 
$2y(1-y)q^2$ term of  ${\cal N}^\mu_{2a}(\ell)$ in (24) which 
cannot cancel the  
$x\,M_{\W}^2$ in ${\cal D}_2(\ell)$ and  gives the dominant 
$\log\,r_\ell$ behavior, reflecting mass singularities (or 
infrared divergences) 
of loop integrals.  
Except this $\log\,r_\ell$, {\it all other terms} of the six vertices 
are negligibly small because they 
are strongly damped by powers of $r_\ell$, or 
$r^n_\ell \, \log\,r_\ell$ where $ n > 0, r_\ell <10^{-3}$. Indeed,
 the four diagrams of Fig.1 are  
strongly damped since infrared-like divergence cannot occur:  
the $x=1$ integration limit of the $(1-x) M_{\W}^2$ in the 
denominator ${\cal D}_1(\ell)$ is systematically canceled by the 
$1-x$ coming from the integration over the $\,y\,$ variable. Also Fig.2b is damped by 
$r_\ell \log\,r_\ell$, due to the $\Phi$-fermions coupling. 

Explicit $x,y$ integrations of all the six vertices 
$\Gamma^\mu_{1a-d}(\ell),\Gamma^\mu_{2a,b}(\ell)$  
confirm these features. The $\log\,r_\ell$ infrared-like divergence, which arises 
when there are two massless ($r_\ell =0$) fermion propagators in 
the loop, has been noticed a 
long time ago in the neutrino charge radius computations$^6$. 

This leading $q^2\log\,r_\ell$ term of Fig.2a  in the 
$\nu_H$-$\nu_L$-$\gamma^*$ vertex  cancels the photon 
propagator $1/q^2$ in Fig.4 and yields  an   effective
 local four-fermion coupling proportional to $G_F$. Thus the 
radiative correction  to the $\nu_H \to \nu_L +$ e$^+$ + e$^-$ 
tree amplitude is found to be 
\begin{equation}
{\cal A}_{rad} = {G_F \over \sqrt{2}} \;{e^2\over 24 \pi^2}
 \Bigl [\sum_{\ell} U^*_{H \ell} U_{L \ell}\;\log r_\ell \Bigr]
 \;\overline{u} (p) \gamma^\mu \Gm  
u(P) \;\overline{u}(k_{-}) \gamma_\mu  v(k_{+})\;\;,
\end{equation}
which can be put in a form similar to ${\cal A}_{tree}$ in (33):
\begin{equation}
{\cal A}_{rad} = {G_F \over \sqrt{2}} \;{e^2\over 24 \pi^2}
 \Bigl [\sum_{\L=\mu,\tau} U^*_{H \L} U_{L \L}\;\log {m^2_{\L} \over m_e^2} 
 \Bigr] \;\overline{u} (p) \gamma^\mu \Gm  
u(P) \;\overline{u}(k_{-}) \gamma_\mu  v(k_{+})\;\;.
\end{equation}
The  sum ${\cal A}_{tree}+{\cal A}_{rad}\equiv {\cal B} $ is now easy to manipulate when we take the interference between ${\cal A}_{tree}$ and  ${\cal A}_{rad}$ in $|{\cal B}|^2 $ for the decay rate. Thus
\begin{equation}
{\cal B} = {G_F \over \sqrt{2}}  \;\overline{u} (p) \gamma^\mu \Gm  
u(P) \;\overline{u}(k_{-}) \gamma_\mu (g_V-g_A\gamma_5)  v(k_{+})\;\;,
\end{equation}
 with
\begin{equation}
g_V =  \sum_{\L=\mu,\tau} U^*_{H \L} U_{L \L} \Bigl(1+ {\alpha\over 3\pi}\log {m_{\L}\over m_e}  \Bigr) \;\,, \;g_A =  \sum_{\L=\mu,\tau} U^*_{H \L} U_{L \L}\;.
\end{equation}
From the amplitude ${\cal B}$, we compute$^7$ the decay rate $\Gamma_1 \equiv \Gamma(\nu_H \to \nu_L +$ e$^+$ + e$^-$)  and find 
\begin{equation}
{\D\Gamma_1\over \D q^2 }  ={G_F^2\over 192\pi^3}  {\sqrt {q^2(q^2-4m_e^2)}\over q^4 M^3} (M^2-q^2)^2
\Bigl\{ (g_V^2+g_A^2) \bigl[ q^2(M^2+2q^2) +2m_e^2 (M^2 -q^2) \bigr] +6 m_e^2 q^2 (g_V^2-g_A^2)\Bigr\} \;,
\end{equation}
from which we get 
\begin{equation}
\Gamma_1 = \int_{4m_e^2}^{M^2} \D q^2 {\D\Gamma_1\over \D q^2 } =  {G_F^2 M^5\over 192\pi^3}\Bigl\{ {g_V^2+g_A^2\over 2} G(x) + (g_V^2-g_A^2)  H(x)\Bigr\}\;, 
\end{equation}
where $x= m_e^2/M^2$, and $G(x), H(x)$ are the phase-space functions given by
\begin{equation}
G(x)= \left[ 1- 14 x - 2 x^2 -12 x^3\right] \sqrt{1-4x} + 24\, 
x^2\,(1-x^2) \log {1+\sqrt{1-4x}\over 1-\sqrt{1-4x}}\;,
\end{equation}
\begin{equation}
H(x)= 2x(1-x)(1+6x)\sqrt{1-4x} + 12x^2 (2x-1-2x^2) \log {1+\sqrt{1-4x}\over 1-\sqrt{1-4x}}\;.
\end{equation}
To this leading logarithm radiative correction expressed by $ \approx \alpha \log r$ in 
(37)-(39), we may also add the nonleading (simple $\alpha$ without $\log r$) electromagnetic  correction to the e$^+$-e$^-$ pair. This nonleading QED correction can be obtained from the one-loop QCD correction to the well known e$^+$+e$^- \to$ quark-pair cross-section, or the $\tau \to \nu_\tau + $ quark-pair decay rate which is already computed in the literature$^2$. The only appropriate change is the substitution $(4/3)\alpha_s \leftrightarrow \alpha$,  because when going from QCD to QED, we replace a gluon with a photon and the quark-pair with the 
electron-pair, the QCD vertex $\I g_s \gamma^\mu \lambda_j/2$ is replaced with the QED vertex $\I e\gamma^\mu$ and the factor $4/3$ comes from $4/3  = (1/4)\sum_j \lambda_j\lambda_j$, where  $\lambda_j$ are the eight Gell-Mann matrices of the color $SU_c(3)$ group. 

This QED nonleading corrected rate $\Gamma_2$ is
 \begin{equation}
\Gamma_2 = {G_F^2 M^5\over 192\pi^3} \left({3\alpha\over 4\pi}\right) G(x) 
K(x,x)\;;
\end{equation}
the function $K(x,x)$ is tabulated in Table 14.1 of reference 2. We emphasize that $K(x,x)$ is a spectacular increasing function of $x$, acting in opposite direction to the decreasing phase-space function  $G(x)$.
As announced, the   $\nu_H\to \nu_L$ + e$^+$ + e$^-$ decay
rate  given by the sum  $\Gamma_1+\Gamma_2$ in (39)--(42) largely dominates the $\nu_H \to \nu_L +\gamma $ in (30). Finally we note that the virtual weak neutral Z boson replacing the virtual photon in Fig.4 also contributes to $\nu_H\to \nu_L$ + e$^+$ + e$^-$. However it can be safely discarded, being strongly damped by $q^2/M_Z^2$ due to the Z propagator.
 
{\bf Conclusions}

The recent observation by the Super-Kamiokande collaboration of a clear
up--down $\nu_\mu$ asymmetry in atmospheric neutrinos is strongly
suggestive of $\nu_\mu \to \nu_{{\rm X}}$ oscillations, where $\nu_{{\rm
X}}$ may be identified with $\nu_\tau$ or even possibly a sterile neutrino.
These results have many important implications in elementary particle  physics and astrophysics$^8$. In particular,
neutrino oscillations mean that neutrinos have a non-vanishing mass, which,
according to the new data, may be at least as heavy as $5\times 10^{-2}\,$
eV. If a neutrino $\nu_{{\rm H}}$ has indeed a mass, it may not be stable and could decay into a lighter neutrino,
$\nu_{{\rm L}}$, through a cross-family electroweak coupling. We have
studied two such decay modes, $\nu_{{\rm H}} \to \nu_{{\rm L}} +\gamma$
and $\nu_{{\rm H}} \to \nu_{{\rm L}}+ {\rm e}^+ + {\rm e}^-$, and found
that the latter,  which, in contrast to the former,  arises at  the
tree level and gets largely enhanced by radiative corrections, is by far the dominant process
and may therefore be detectable provided that $\nu_{{\rm H}}$ has a mass
$>2\,m_{{\rm e}}$. A positive evidence for such decay modes would give
a clear signal of the onset of `new physics'.

\vskip1mm
{{\bf EXERCISES}\hskip6mm 1- Show that the $\xi$-dependences of  Fig.2a and Fig.2b amplitudes
exactly cancel out, leaving their sum $\xi$-independent.

\hskip6mm2- Explain qualitatively why the lifetimes jump from $10^{-2}$ year to 
$10^{44}$ years when the neutrino mass changes from 1.1 MeV to $5\times 10^{-2}$ eV.

{\bf Figure Captions}  :  \normalsize 


Figures 1--2 : One-loop $\nu_H \to \nu_L +\gamma$ in the renormalizable 
$R_\xi$-gauge.

Figure 3 : Tree diagram  $\nu_H \to \nu_L + $e$^+$ + e$^-$

Figure 4 : Leading radiative corrections to $\nu_H \to \nu_L +$e$^+$ + e$^-$




\begin{thebibliography}{99}
\bibitem{7} The calculation of the $F\to f_1+f_2 +\overline{f}_3$
 decay rate is explained for instance in ref. 2, chapter 13, where $F, f_i$ stand for fermions
\end{thebibliography}
\end{document}